\documentclass{Interspeech2024}
\usepackage{graphicx}
\usepackage{multirow}
\usepackage{hyperref}
\usepackage{url}

\interspeechcameraready

\title{Towards Expressive Zero-Shot Speech Synthesis with Hierarchical Prosody Modeling}

\name[affiliation={1}]{Yuepeng}{Jiang}
\name[affiliation={1}]{Tao}{Li}
\name[affiliation={2}]{Fengyu}{Yang}
\name[affiliation={1*}]{Lei}{Xie}
\name[affiliation={2}]{Meng}{Meng}
\name[affiliation={2}]{Yujun}{Wang}

\address{
  $^1$Audio, Speech and Language Processing Group (ASLP@NPU), School of Software,\\
  Northwestern Polytechnical University, Xi’an, China\\
  $^2$Xiaomi AI Lab, Beijing, China\vspace{-2pt}}
\email{jiangyp@mail.nwpu.edu.cn, lxie@nwpu.edu.cn\thanks{* Corresponding author.}}

\keywords{speech synthesis, zero-shot, prosody modeling, denoising diffusion probabilistic model}

\begin{document}

\maketitle

\begin{abstract}
Recent research in zero-shot speech synthesis has made significant progress in speaker similarity.
However, current efforts focus on timbre generalization rather than prosody modeling, which results in limited naturalness and expressiveness.
To address this, we introduce a novel speech synthesis model trained on large-scale datasets, including both timbre and hierarchical prosody modeling.
As timbre is a global attribute closely linked to expressiveness, we adopt a global vector to model speaker timbre while guiding prosody modeling.
Besides, given that prosody contains both global consistency and local variations, we introduce a diffusion model as the pitch predictor and employ a prosody adaptor to model prosody hierarchically, further enhancing the prosody quality of the synthesized speech.
Experimental results show that our model not only maintains comparable timbre quality to the baseline but also exhibits better naturalness and expressiveness. 
The synthesized samples can be found at: \url{https://rxy-j.github.io/HPMD-TTS/}

\end{abstract}

\section{Introduction}

The aim of text-to-speech (TTS) synthesis is to convert text into human-like speech.
With the development of neural networks and deep learning, the naturalness and intelligibility of the synthesized speech for speakers \textit{seen} in the training corpus have been promoted to a human-like level~\cite{tacotron, fastspeech, gradtts, vits, delightfultts}.
However, such systems can only generate voice for speaker-limited recording-studio datasets, which limits the diversity of generated speaker timbre, prosody, and style~\cite{naturalspeech2}.
Therefore, how to generate speech for any speaker who was \textit{unseen} during the TTS model training, solely relying on a single acoustic reference of that speaker, i.e. zero-shot speech synthesis, is gaining growing attention recently~\cite{7789, jia2018transfer, yourtts}.


In zero-shot speech synthesis, the primary challenge lies in enhancing the robustness and generalizability of speaker timbre modeling~\cite{speakermodeling}, thereby guaranteeing a precise reproduction of the unseen target speaker's timbre in synthesized speech.
In prior studies, a commonly employed speaker modeling approach has been to leverage a speaker encoder to extract speaker embeddings from the reference speech, subsequently as a speaker timbre control condition in the acoustic model.
Based on training methods, the current speaker encoders can be roughly classified into jointly trained~\cite{10101112, 10111212} and pre-trained methods~\cite{7789, 7889, 7899}.
For jointly trained methods, researchers focus on training an encoder network jointly with acoustic models, like using unsupervised latent embedding technologies~\cite{metastyle, anyspeaker}, such as global speaker embeddings (GSEs) ~\cite{speakerembedding} or variational autoencoder (VAE) ~\cite{hgmcss}.
However, these methods are plagued by unsatisfactory generalization performance due to training data's scale limitations, resulting in poor speaker similarity.
Pre-trained methods draw on transfer learning, where speaker encoders are independently trained based on other tasks, e.g., speaker verification or speech recognition.
In this case, the speaker encoder utilizes large amounts of data for training and can provide more robust representations of speaker characteristics. 
ZSM-SS~\cite{zsmss} constructs a speaker encoder based on Wav2vec 2.0~\cite{wav2vec2} and achieves good performance in speaker similarity. 
YourTTS~\cite{yourtts} adopts an encoder pre-trained based on speaker recognition and designs a speaker consistency loss function to optimize model training, achieving state-of-the-art results in zero-shot testing on the VCTK dataset. 
Despite these advances, the generalization capability of these methods is still insufficient, and the naturalness and timbre similarity are still not ideal in the case of synthesizing speech for unseen speakers.

Recently, zero-shot TTS methods~\cite{valle, naturalspeech2, megatts} that rely on large-scale corpora have gained increasing attention.
The diversity of data available in such corpora has significantly enhanced the generalization capabilities of TTS models, enabling them to capture the diverse speaker timbre, prosody, and style, effectively remarkable speaker similarity of unseen speakers.
Based on their data representation methods, these methods can be classified into two types: discrete tokens and continuous vectors. 
VALL-E~\cite{valle} uses a neural codec based on Residual Vector Quantization (RVQ) to quantize speech into tokens, taking both text and audio tokens as input, while a language model is used to generate the target speech tokens autoregressively. 
SPEAR-TTS~\cite{speartts}, modifies AudioLM~\cite{audiolm} and achieves a stepwise transformation from text to semantic tokens and then to acoustic tokens. 
NaturalSpeech2~\cite{naturalspeech2} uses a neural codec based on RVQ to encode audio into continuous vectors as its data representation. Then, a latent diffusion model~\cite{latentdiffusion} is adopted to predict these continuous vectors.
However, these methods neglect explicit modeling of prosody, which could result in insufficient naturalness and expressiveness of the generated speech, especially when the reference speech is already highly expressive and varied in prosody.


\begin{figure*}[!htbp]
    \centering
    \includegraphics[scale=0.65]{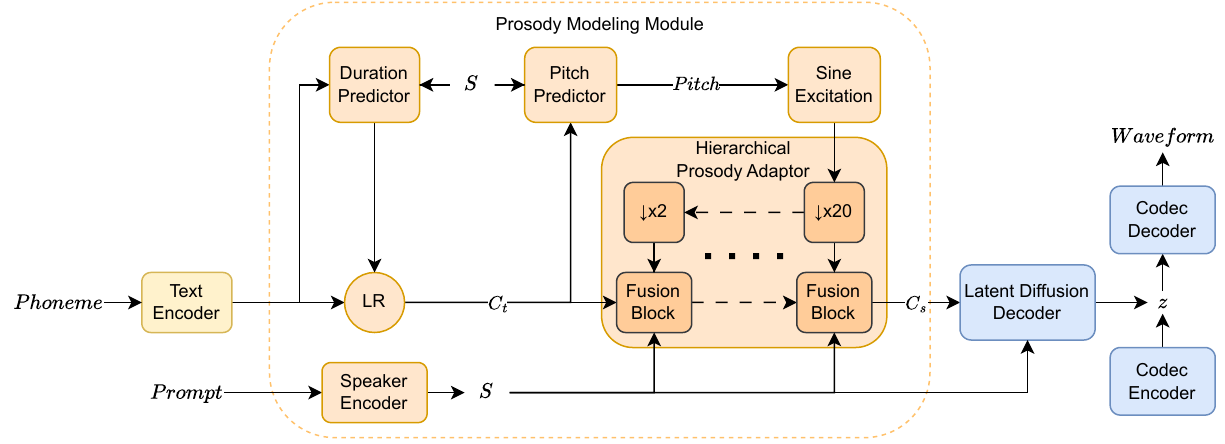}
    \caption{Architecture overview of the proposed model}
    \label{fig:system_overview}
\end{figure*}

To harness the diversity of large-scale corpora and simultaneously enhance the prosody modeling capabilities of the model, we introduce a novel zero-shot speech synthesis model that incorporates a diffusion model-based prosody predictor and a hierarchical prosody adaptor.
First, considering that speaker timbre is a global attribute that does not vary over time, we adopt a speaker encoder to extract global speaker embedding. 
As quantized audio excessively discards speech details, it fails to guide speaker timbre restoration. Therefore, we employ mel spectrograms as inputs to the speaker encoder.
Second, we employ a diffusion model as a pitch predictor to match speech prosody diversity by leveraging its natural advantage in generating content diversity.
Finally, since prosody shows both global consistency and local variations, we employ a prosody adaptor to model prosody hierarchically, such as frame-level, phoneme-level, and word-level, to improve the prosody performance of synthesized speech. 
Experimental results demonstrate that our model achieves better naturalness while maintaining speaker similarity compared to the comparison models. 
The results also prove that utilizing a hierarchical modeling approach significantly improves prosody details and jitter issues compared to directly adding prosody information to intermediate variables.

\section{Proposed Approach}


\subsection{Overview}

As shown in Figure~\ref{fig:system_overview}, the proposed model consists of a text encoder, a prosody modeling module, and a latent diffusion decoder. 
The prosody modeling module consists of a speaker encoder, a diffusion-based pitch predictor, a duration predictor, and a hierarchical prosody adaptor.
The diffusion-based pitch predictor uses the text encoder output $C_t$ and the global speaker representation $S$ as conditions to predict the fundamental frequency. 
The hierarchical prosody adaptor uses the fundamental frequency for hierarchical prosody modeling. Meanwhile, guided by the global speaker representation $S$, it performs speaker timbre modeling, transforming the text encoder output $C_t$ into a content representation containing prosody and timbre information $C_s$. 
Then, the latent diffusion decoder, conditioned on $C_s$ and $S$, generates the latent vector $z$ and further enhances the timbre information in $z$. 
Finally, a waveform is reconstructed from $z$ by the codec decoder.

\subsection{Pitch Predictor Based on Diffusion}

Given the diversity of speech prosodic patterns, we adopted a diffusion model as a predictor to leverage its natural advantage in generating content diversity.
The denoising diffusion probabilistic model (DDPM) is a generative model that uses a predefined forward process to gradually transform data into pure noise. 
Subsequently, the model learns the reverse operation of this process, through the denoising process, it gradually reconstructs the original data from the noise.
Its forward process can be represented using the following equation:
\begin{equation}
    q\left(\mathbf{x}_t \mid \mathbf{x}_{t-1}\right):=\mathcal{N}\left(\mathbf{x}_t ; \sqrt{1-\beta_t} \mathbf{x}_{t-1}, \beta_t \mathbf{I}\right).
\end{equation}

At each step of the diffusion process, a very tiny Gaussian noise is added to $x_{t-1}$ at the current time $t-1$ to obtain $x_t$. 
This Gaussian noise is determined by variance schedule $\beta=\{\beta_1, \cdots, \beta_T\}$. 
The complete forward process can be described as follows:
\begin{equation}
    q\left(\mathbf{x}_{1: T} \mid \mathbf{x}_0\right):=\prod_{t=1}^T q\left(\mathbf{x}_t \mid \mathbf{x}_{t-1}\right),
\end{equation}
where $T$ is the number of steps when the diffusion process terminates. 
When $T$ is large enough, $x_T$ can be regarded as random noise following the Gaussian distribution $N(0, I)$. 
Therefore, we can let the reverse process start at $x_T \sim N(0, I)$. 
For each step of the reverse process, it can be described by the following equation:
\begin{equation}
    p_\theta\left(\mathbf{x}_{t-1} \mid \mathbf{x}_t\right):=\mathcal{N}\left(\mathbf{x}_{t-1} ; \boldsymbol{\mu}_\theta\left(\mathbf{x}_t, t\right), \sigma_t^2 \mathbf{I}\right),
\end{equation}
where $\sigma^2=\frac{1-\bar{\alpha}_{t-1}}{1-\bar{\alpha}_t} \beta_t$ $\alpha_t:=1-\beta_t$ and $\bar{\alpha}:=\prod_{s=1}^t \alpha_s$. $\theta$ is a learnable parameter. 
In general, there is no direct way to obtain the forward process $q(x_t|x_{t-1})$ corresponding to the reverse process at time $t$. 
Therefore, we use a neural network to predict $\theta$, allowing it to approximate the forward process at time $t$.
To train parameter $\theta$, we minimize the variational bound of negative log-likelihood.~\cite{ddpm} has simplified it:
\begin{equation}
L(\theta):=\mathbb{E}_{t, \mathbf{x}_0, \boldsymbol{\epsilon}}\left[\Vert\boldsymbol{\epsilon}-\boldsymbol{\epsilon}_\theta\left(\sqrt{\bar{\alpha}_t} \mathbf{x}_0+\sqrt{1-\bar{\alpha}_t} \boldsymbol{\epsilon}, t\right)\Vert^2\right],
\end{equation}
where sample $\epsilon \sim N (0, I)$ and $\epsilon(\cdot)$ is the outputs of the neural network.
The complete proof of denoising diffusion probabilistic models can be found in~\cite{ddpm}.

For the model we are using, the denoiser predicts $\theta$ by using $C_t$ and $S$ as conditions. Therefore, in this model, the actual denoising process is described by the following equation:
\begin{equation}
    \mathbf{x}_{t-1}=\frac{1}{\sqrt{\alpha_t}}\left(\mathbf{x}_t-\frac{1-\alpha_t}{\sqrt{1-\bar{\alpha}_t}} \boldsymbol{\epsilon}_\theta\left(\mathbf{x}_t, C_t, s, t\right)\right)+\sigma_t \mathbf{z}.
\end{equation}


Moreover, we apply gradient clipping to its inputs. 
Therefore, the predictor in this model is a completely independent module.

\begin{table*}[!htbp]
    \centering
    \caption{Evaluations for different models  with 95\% confidence intervals.}
    \label{eval}
        \begin{tabular}{ccccc|cccc}
        \toprule[\heavyrulewidth]
        \multirow{2}{*}{\textbf{Method}} & \multicolumn{4}{c|}{Emotional}      & \multicolumn{4}{c}{Overall}                           \\ \cline{2-9}
                          & \textbf{NMOS $\uparrow$} & \textbf{SMOS $\uparrow$} & \textbf{CER(\%) $\downarrow$} & \textbf{SECS $\uparrow$} & \textbf{NMOS $\uparrow$} & \textbf{SMOS $\uparrow$} & \textbf{CER(\%) $\downarrow$}& \textbf{SECS $\uparrow$}  \\ \midrule[\heavyrulewidth]
        Proposed          &  \textbf{3.65}$\pm$0.14    &   \textbf{3.50}$\pm$0.15  &  \textbf{6.35}    &  0.818   &  \textbf{3.62}$\pm$0.10  &   3.54$\pm$0.11          &  \textbf{5.99}    &  \textbf{0.829}   \\ 
        NS2               &  3.50$\pm$0.14    &   3.30$\pm$0.17           &  7.43             &  0.807            &  3.54$\pm$0.13           &   3.36$\pm$0.15          &  6.56             &  0.815            \\ 
        VALL-E            &  3.52$\pm$0.16    &   3.49$\pm$0.22           &  13.07            &  \textbf{0.820}   &  3.60$\pm$0.13           &   \textbf{3.57}$\pm$0.17 &  10.18            &  0.826            \\ 
        \bottomrule[\heavyrulewidth]
        \end{tabular}
\vspace{-15pt}
\end{table*}

\subsection{Hierarchical Prosody Adaptor}

Considering that the global consistency and local variability in prosody, we adopt a prosody adaptor to model the prosody in a hierarchical manner.
The detailed structure of the hierarchical prosody adaptor is illustrated in Figure~\ref{fig:system_overview}. 
This module takes the content information $C_t$, frame-level pitch, and speaker embedding $S$ as inputs and outputs $C_s$ containing prosody, timbre, and content information. 
Inspired by~\cite{dspgan}\cite{svc}, we utilize sample-level sinusoidal excitation signals to further represent pitch information. 
Firstly, we need to expand the frame-level f0 to the sample-level f0 $f_0[n]$ through linear interpolation. 
Subsequently, the corresponding sine excitation signals $p[n]$ can be generated by the following two formulas:

\begin{equation}
    p[n]= \begin{cases}\sum_{k=1}^K \sin \left(\frac{2 \pi k \sum_{i=0}^n f_0[i]}{f_s}\right), & f_0[n]>0 \\ 0, & f_0[n]=0\end{cases}
\end{equation}

\begin{equation}
    K= min(K_{max}, \left\lfloor\frac{f_s}{2 f_0[n]}\right\rfloor),
\end{equation}
where $k \in \{1, 2, ..., K\}$ and $n$ are the indexes of harmonics and time, respectively. 
$K$ is the largest number of harmonics that can be achieved corresponding to the $f_0[n]$. $f_s$ is the audio sampling rate. 
In this paper, we set $K_{max}$ to 200, which limits the largest harmonics index to 200.

By downsampling the sinusoidal excitation signal multiple times, we can obtain information at different scales. 
Unlike the method proposed in~\cite{dspgan}, it is not required for the length of the sinusoidal signal to match the length of the content information in our module. 
Therefore, we increase the downsampling factor to ensure that the scale difference between the two signals is maximized. 
During the process of multiple downsampling, the scale of the signal can approximate frame-level, phoneme-level, and word-level, respectively. 
This provides hierarchical prosody information and facilitates modeling at different scales.

In the system we propose, there are a total of four downsampling layers and corresponding modeling units. 
We set downsampling factors to [20, 10, 10, 2], respectively. 
During the data preprocessing process, the hop size is set to 200 when extracting the fundamental frequency, which means that the data needs to be downsampled by 200 times from the sample-level to the frame-level. 
To prevent the size of a single downsampling layer from being too large, we achieve 200 times downsampling through the first two downsampling layers.
According to research~\cite{speechrate}, the average speaking rate in Chinese is approximately 150 to 200 characters per minute. 
Considering that our phoneme set comprises an average of two phonemes per character, and accounting for factors such as speech pauses, the ratio between frame-level and phoneme-level is approximately 10 to 15. 
Therefore, we set the downsampling factor for the third level to 10 and for the fourth level to 2, corresponding to the phoneme-level and word-level, respectively.
Subsequently, we employ cross-attention at each scale to fuse prosody information.

Additionally, to better integrate the global speaker timbre and prosody information $S$, we introduce the SALN module for information injection, ensuring that each prosody modeling unit can observe global and fine-grained prosody information at the same time. 
This approach not only preserves the consistency of global features but also enhances the system's adaptability to different speakers, further strengthening the model's ability in speech modeling.


%

\section{Experiments}

\subsection{Dataset and Preprocessing}

The WenetSpeech4TTS~\cite{wenet4tts} dataset is used in experiments. 
Considering that it is originally a Mandarin corpus designed for speech recognition, we process it to make it applicable for TTS.
The annotated data is re-segmented, then denoised and filtered to exclude speech with multiple speakers. 
The data is then scored using DNSMOS~\cite{dnsmos2021}, and we obtain about 8000 hours of data with scores above 3.6 and about 4000 hours with scores above 3.8, denoted as Large (\textit{L}) and Medium (\textit{M}), respectively.
These two sets of data are used for subsequent model training. 
The data is resampled to 16k and the loudness is normalized.


To validate the zero-shot voice cloning capability, prosody modeling ability, and expressiveness of our proposed model, we select 15 speakers for evaluation. 
Specifically, these speakers are chosen from AISHELL3\footnote{\url{https://www.aishelltech.com/aishell_3}}, an internal emotional dataset~\cite{Li2021ControllableCE}, and WenetSpeech's evaluation set, with 5 speakers selected from each source.
The internal emotion dataset comprises six typical emotions i.e., happy, anger, sad, fear, surprise, and disgust. 
By utilizing speakers selected from this dataset, we can evaluate the naturalness performance of the proposed model when presented with highly expressive references.

We process the text using pypinyin\footnote{\url{https://github.com/mozillazg/python-pinyin}}, extracting phonemes and tones as model inputs.  
In addition, the mel spectrogram is extracted from the original waveforms and randomly samples 1 to 3 seconds of segments as a prompt based on the audio length. 
The fundamental frequency is extracted using PyWORLD\footnote{\url{https://github.com/JeremyCCHsu/Python-Wrapper-for-World-Vocoder}} toolkit, and we apply interpolation for smoothing. 
We utilize Encodec~\cite{encodec} trained on the WenetSpeech dataset as the model's acoustic codec to extract $z$ for training. 
While maintaining the codebook size at 1024, we increase the codebook dimension from 128 to 256.
We use a Kaldi\footnote{\url{https://github.com/kaldi-asr/kaldi}} model trained on a 2000-hour subset of the set \textit{L} to obtain external duration information.

\subsection{Training}

For the sinusoidal excitation signal extraction, we set $K_{max}$ to 200. 
The diffusion steps for the pitch predictor and latent diffusion decoder are set to 100 and 200, respectively. 
We utilize AdamW optimizer for training, with $\beta_1$ set to 0.8, $\beta_2$ set to 0.99, and weight decay $\lambda$ set to 0.01. 
The initial learning rate is set to 1e-4. Our model is trained up to 600k training steps on 4 NVIDIA A6000 GPUs with a batch size of 16. 
We first train the model on the set \textit{L} and then fine-tune it on the set \textit{M}.

\vspace{-2pt}
\subsection{Baseline}

We compare our proposed model with NaturalSpeech2 (NS2) ~\cite{naturalspeech2} and VALL-E~\cite{valle}. We use the NS2 and VALL-E code provided by the Amphion~\cite{amphion} toolkit to build these models, using their default configurations. The training strategy for both models is similar to ours.

\vspace{-0.25cm}
\section{Results}

\subsection{Objective Evaluation}

We evaluate the model performance on a reserved testset of 150 sentences using Character Error Rate (CER) and speaker embedding cosine similarity (SECS). 
To reduce the error rate caused by homophone transcription mistakes, we carefully select texts that do not contain proper nouns, names, and other error-prone terms to compose this testset.
We use Paraformer~\cite{paraformer} pre-trained on 60,000 hours of Mandarin Chinese to transcribe the generated speech and calculate CER. 
For SECS, we use Resemblyzer~\footnote{\url{https://github.com/resemble-ai/Resemblyzer}} to extract speaker embedding and calculate cosine similarity using the function build in pytorch.
For Paraformer and Resemblyzer, we utilize the official code and pre-trained checkpoints. Objective test results are shown in the table~\ref{eval}.
For comparison, the test results of the speakers from the emotional dataset are listed separately.

As shown in table~\ref{eval}, our model achieves the lowest CER, further reducing it by approximately 0.5\% compared to NS2, which is also a non-autoregressive model. 
The NS2 training process involves extracting prompts from the original audio based on phoneme boundaries and then synthesizing the remaining parts, such a procedure is extremely dependent on the accuracy of external duration information. 
Specifically, the data quality of WenetSpeech is relatively lower than recording-studio benchmarking datasets (e.g., DiDiSpeech~\cite{didispeech}), resulting in lower duration accuracy of WenetSpeech data obtained through Kaldi, which could be the main factor affecting the CER of NS2.
Besides, VALL-E has significantly higher CER due to autoregressive model instability, misreadings, and incorrect pauses.
For the SECS test, there were no significant differences in the scores of the three models, whether analyzing overall performance or the emotional dataset.


\subsection{Subjective Evaluation}

We conduct Mean Opinion Score (MOS) tests to evaluate the speaker similarity and naturalness of synthesized speech, denoted as SMOS and NMOS, respectively. 
Each MOS test involves 20 listeners who rate the speech on a scale from 1 to 5 based on subjective perception. 
The results of the subjective test are also shown in the table~\ref{eval}. 


Through comparison, it can be found that the proposed model outperforms NS2 and VALL-E in both speaker similarity and naturalness. 
Specifically, when dealing with speakers from the emotional dataset, the naturalness of the proposed model does not significantly decline, while VALL-E and NS2 show varying degrees of naturalness degradation, especially VALL-E, which shows the most notable decline.
Furthermore, the listening tests also show that the audio quality of VALL-E and NS2 decreases significantly when dealing with speakers from the emotional dataset, which may be an important reason for their poor performance in speaker similarity and naturalness.

\begin{table}[h]
    \centering
    \caption{Results of ablation studies.}
    \label{as}
    \resizebox{\linewidth}{!}{
        \begin{tabular}{lc}
            \toprule
            \textbf{Model}                          & \textbf{NMOS ($\uparrow$)} \\ \midrule
            Proposed                                &  3.62$\pm$0.10             \\    
            ~- pitch predictor based on diffusion   &  3.58$\pm$0.07             \\ 
            ~- hierarchical prosody modeling         &  3.49$\pm$0.16             \\ \bottomrule
        \end{tabular}
    }
\end{table}

\vspace{-0.5cm}
\subsection{Ablation Study}

To validate the effectiveness of our model structure, we carry out a series of ablation studies. 
We conduct ablation experiments on the pitch predictor and hierarchical prosody modeling. 
In the ablation study for the pitch predictor, we replace our model's predictor with the pitch predictor from the FastSpeech~\cite{fastspeech}. 
For the hierarchical prosody modeling ablation, we remove the related structures. We directly encode frame-level fundamental frequency information and then add it to $C_t$.

\begin{figure}
    \centering
    \includegraphics[scale=0.59]{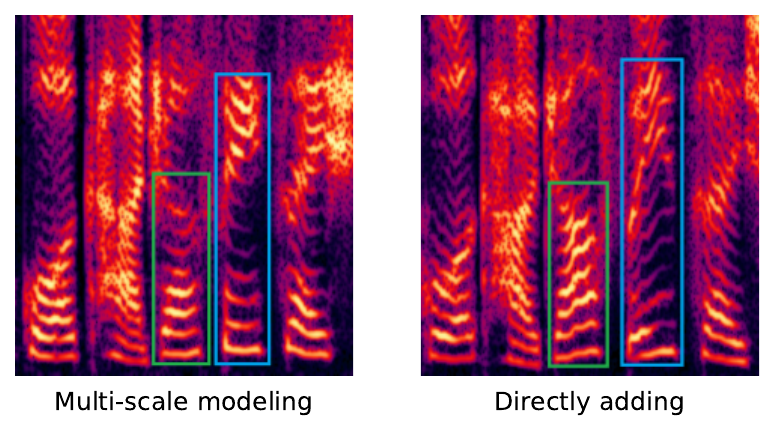} 
    \caption{The spectrograms of synthesized samples in hierarchical prosody modeling ablation.}
    \label{fig:jitter_compare}
\vspace{-0.5cm}
\end{figure}

The experimental results show that removing the fundamental frequency predictor leads to a slight decrease in naturalness, indicating that the introduction of diffusion does indeed play a positive role in enhancing prosody diversity. 
The significant decrease in naturalness after removing hierarchical prosody modeling highlights the importance of implementing prosody modeling at different scales. 
Furthermore, in comparative listening tests, we find that the direct addition of fundamental frequency information to $C_t$ causes noticeable pronunciation jitter, shown in Figure~\ref{fig:jitter_compare}, which makes the speech sound unnatural.

\vspace{-4pt}
\section{Conclusions}


To leverage the data diversity of large-scale corpora and simultaneously enhance the prosody modeling performance, this paper proposes a novel zero-shot speech synthesis model that incorporates a diffusion model-based prosody predictor and a hierarchical prosody adaptor. 
Specifically, we employ global vectors to represent speaker timbre and utilize them as a guide for prosody modeling as the expressiveness is correlated to the speaker.
Considering that prosody is reflected in both global consistency and local variations, a diffusion model is introduced as the pitch predictor while a prosody adaptor is adopted to model hierarchical prosody variance, thus enhancing the prosody performance of the synthesized speech.
Experimental results show that our model not only maintains comparable timbre quality to the baseline but also exhibits better naturalness and expressiveness.

\bibliographystyle{IEEEtran}
\bibliography{mybib}

\end{document}